%% file: main.tex
\documentclass[11pt,prb,reprint,superscriptaddress,citeautoscript]{revtex4-1}
\usepackage[caption=false]{subfig}
\usepackage[version=4]{mhchem}
\usepackage{graphicx,color}
\usepackage[usenames,dvipsnames,svgnaes,table]{xcolor}
\usepackage{bm}
\usepackage{amsmath}
\usepackage{amssymb}
\usepackage{mathtools}
\usepackage{siunitx}
\usepackage[super]{nth}
\usepackage{upgreek}
\usepackage{multirow}
\usepackage{array}
\usepackage{color}
\usepackage{longtable}
\usepackage{xfrac}
\usepackage{hyperref}
\usepackage{svg}
\usepackage{graphicx, type1cm, lettrine}

\hypersetup{colorlinks=true,
            linkcolor=red,
            citecolor=blue,
            urlcolor=orange}

\newcolumntype{K}[1]{>{\centering\arraybackslash}p{#1}}
\setcitestyle{numbers,super}

\usepackage{ amssymb }
\usepackage{ amsmath }
\usepackage{ amsfonts}
\usepackage{ dsfont }
\usepackage{ amsthm }
\usepackage{algpseudocode, algorithm}
\usepackage{bbm}
\usepackage{enumerate}
\usepackage{hyperref}
\usepackage{dsfont}
\usepackage{bm}

\DeclareMathOperator*{\argmin}{arg\,min}

\AtBeginDocument{\usepackage{booktabs}}

\begin{document}

\title{Probabilistic Phase Labeling and Lattice Refinement for Autonomous Material Research}

\author{Ming-Chiang Chang}
\email{mc2663@cornell.com}
\affiliation
{Department of Materials Science and Engineering, Cornell University, Ithaca, NY 14853, United States}

\author{Sebastian Ament}
\affiliation
{Department of Computer Science, Cornell University, Ithaca, NY 14853, United States}

\author{Maximilian Amsler}

\affiliation{Department of Materials Science and Engineering, Cornell University, Ithaca, NY 14853, United States}

\author{Duncan R. Sutherland}
\affiliation{Department of Materials Science and Engineering, Cornell University, Ithaca, NY 14853, United States}

\author{Lan Zhou}
\affiliation
{Division of Engineering and Applied Science and Liquid Sunlight Alliance, California Institute of Technology, Pasadena, CA, USA}

\author{John M. Gregoire}
\affiliation
{Division of Engineering and Applied Science and Liquid Sunlight Alliance, California Institute of Technology, Pasadena, CA, USA}

\author{Carla P. Gomes}
\affiliation
{Department of Computer Science, Cornell University, Ithaca, NY 14853, United States}

\author{R. Bruce van Dover}
\affiliation{Department of Materials Science and Engineering, Cornell University, Ithaca, NY 14853, United States}

\author{Michael O. Thompson}
\email{mot1@cornell.edu}
\affiliation{Department of Materials Science and Engineering, Cornell University, Ithaca, NY 14853, United States}

\begin{abstract}
X-ray diffraction (XRD) is an essential technique to determine a material's crystal structure in high-throughput experimentation, and has recently been incorporated in artificially intelligent agents in autonomous scientific discovery processes. However, rapid, automated and reliable analysis method of XRD data matching the incoming data rate remains a major challenge. To address these issues, we present CrystalShift, an efficient algorithm for probabilistic XRD phase labeling that employs symmetry-constrained pseudo-refinement optimization, best-first tree search, and Bayesian model comparison to estimate probabilities for phase combinations without requiring phase space information or training. We demonstrate that CrystalShift provides robust probability estimates, outperforming existing methods on synthetic and experimental datasets, and can be readily integrated into high-throughput experimental workflows. In addition to efficient phase-mapping, CrystalShift offers quantitative insights into materials' structural parameters, which facilitate both expert evaluation and AI-based modeling of the phase space, ultimately accelerating materials identification and discovery.

\end{abstract}

\maketitle

\section{Introduction}
\input{introduction}

\section{Results}
\input{result}

\section{Discussion}
\input{discussion}

\section{Methods} 
\label{sec:methods}
\input{method}

\section{Code Availability}
The codes pertaining to the current study can be found at \\
\href{https://github.com/MingChiangChang/CrystalShift.jl}{https://github.com/MingChiangChang/CrystalShift.jl} and \href{https://github.com/MingChiangChang/CrystalTree.jl}{https://github.com/MingChiangChang/CrystalTree.jl}.

\section{Author Contribution}
M.T., R.D., C.G. and J.G. identified the problem to be solved.
M.C., S.A., J.G. and M.T. conceptualized the algorithm.
M.C. and S.A. developed, design the framework, and implemented it.
M.C. and S.A. designed the tests and M. C. performed the tests.
L.Z. performed material synthesis and characterization for Cr-Fe-V-O film.
D.S. synthesized the Ta-Sn-O thin film. M.C., D.S., M.A. and M.T. performed LSA and XRD of Ta-Sn-O thin-film, and the preliminary data process before feeding to CrystalShift.

\section{Acknowledgments}\label{sec:ack}
The authors acknowledge the Air Force Office of Scientific Research for support under award FA9550-18-1-0136. This material is based on research sponsored by AFRL under agreement number FA8650-19-2-5220. The U.S. Government is authorized to reproduce and distribute reprints for Governmental purposes notwithstanding any copyright notation thereon.
This work was performed in part at the Cornell NanoScale Facility, a member of the National Nanotechnology Coordinated Infrastructure (NNCI), which is supported by the National Science Foundation (Grant NNCI-2025233). MA acknowledges support from the Swiss National Science Foundation (project P4P4P2-180669). This research was conducted with support from the Cornell University Center for Advanced Computing.

\bibliography{references}

\end{document}

%% file: introduction.tex
Synthesizing materials and optimizing material properties to meet requirements for target applications is challenging due to the inherent complexity of composition-structure-property relationships among candidate materials. 
In addition to the chemical space's combinatorial explosion that arises from the consideration of an increasing number of chemical elements, 
the various materials processing conditions introduce additional continuous degrees of freedom, leading to a materials design space that is impossible to tackle with conventional material synthesis methods.
To explore such a high-dimensional space, high-throughput experiment (HTE) methods have been developed to accelerate material research and have been successfully applied for the discovery of various classes of functional materials, such as advanced alloys\citep{guerin2008high,  moorehead2020high}, energy storage materials\citep{otani2007high, matsuda2019high, jonderian2022suite}, and for automated screening in pharmaceutical drug design.\citep{mennen2019evolution, ting2016high} The full potential of HTE is unleashed only when complemented by high-throughput characterization and data analysis methods at matching timescales to avoid workflow bottlenecks and streamline the synthesis-characterization process.\citep{ludwig2019discovery, ament2019multi, clayson2020high}

A key objective in HTE is to establish a better understanding of structure-property relationships, which in turn provides improved rules for advanced materials design. X-ray diffraction (XRD) is particularly well suited to resolve the atomic structure of crystalline phases and, if performed in a high-throughput fashion, to rapidly establish structure-property maps.\citep{cullity2014elements, gregoire2014high} Although many experimental high-throughput XRD frameworks have been developed,\citep{gregoire2014high, lyu2017high} only few analysis methods to extract physical insights from the resulting data at a high rate or even on-the-fly  are available; e.g., traditional techniques such as Rietveld refinement are computationally involved, requires extensive expert knowledge, and is insufficiently robust to match the requirements of HTE.

In addition, material scientists recently started to deploy artificial intelligent (AI) and machine learning techniques to complement and accelerate experimental material discovery efforts.\citep{szymanski2021toward, stach2021autonomous, gregoire2023combinatorial, szymanski2023adaptively} For example, closed-loop experiments based on active learning AI agents have emerged in recent years\citep{dai2020efficient, ament2021autonomous}, which  require no human intervention and can efficiently achieve designated objectives, e.g., mapping material design space with minimal effort, or synthesizing material with desired properties. However, due to the lack of rapid and reliable data analysis method to fully process XRD data for conclusive structural determination, many proposed algorithms operate on reduced quantities such as scalar performance metrics or gradients in spectroscopic signals, which limit the reasoning ability of AI agents.\citep{yuan2018accelerated, kusne2020fly, ament2021autonomous} 
Full structure determination, including composition-dependent lattice parameters, is however central to learning and exploiting composition-structure-property measurements. Establishing an automated XRD analysis framework will enable the development of new autonomous workflows in which AI agents can design synthesis methods to obtain specific atomic arrangements that are conducive to the target properties.

Even for human experts, XRD patterns are notoriously difficult to interpret,
especially if they exhibit complex peak shifting, broadening, and changing peak ratios.
\citep{cullity2014elements, rietveld1969profile} Furthermore, the presence of multiple phases in a single sample gives rise to a mixture of convoluted single-phase XRD patterns, which causes overlapping peaks and poses an additional challenge for accurate phase separation and identification. Since errors in phase labelling can alter the inferred scientific knowledge, and some XRD patterns may be interpreted to contain different phase mixtures,
a labeling algorithm that gives probability estimation of the labels is preferable. Probabilistic labeling and uncertainty estimates are further crucial elements of any robust and efficient AI-based phase space exploration approach.\citep{houlsby2011bayesian, van2020uncertainty}

Because single-phase XRD patterns are much easier to label, a common approach is to process XRD patterns in large batches and leverage source separation tools, such as graph cutting methods and convolutional nonnegative matrix factorization, to separate single phase bases and their corresponding activations, i.e. intensity factors proportional to the fraction of the material that has crystallized into each phase.\citep{kusne2015high, suram2017automated}
However, the conditions that guarantee basis separation cannot always be met,\citep{fu2018identifiability} especially in materials discovery efforts where materials are synthesized from novel compositions that may crystallized into new combinations of phases wherein each phase may be described by lattice constants that differ from previously observed prototypes.

Recently, deep learning methods,\citep{lecun2015deep} in particular convolutional neural network-based methods,\citep{lecun1998gradient} have been applied to solve the multi-phase labeling problem.\citep{szymanski2021probabilistic, maffettone2021crystallography, banko2021deep, chen2021automating, lee2020deep} These methods generally start by creating a training dataset, e.g., by simulating  XRD patterns of phases in crystallographic structure databases like the ICSD \citep{zagorac2019recent} or the Materials Project \citep{jain2013commentary}. This training dataset is at times augmented by simulating the patterns of strained lattices in different direction, by including patterns with different peak widths and, in some cases, by adding  synthetic background signals. This dataset is subsequently used to train a convolutional neural network, creating a phase labelling models that in certain settings outperforms
traditional full pattern matching or correlation methods.\citep{gilmore2004high, hernandez2017using} However, phase coexistence still imposes great difficulties for neural networks trying to separate and identify phases correctly. Some deep learning methods attempt to deal with this spectra separation problem by executing a detect-and-subtract method, i.e., to detect a phase, subtract its signal from the XRD pattern, and repeat this process to iteratively identify the remaining phases until the complete XRD pattern is sufficiently reconstructed.\citep{szymanski2021probabilistic} This requires less training sample but can be vulnerable to experimental noise and strong overlap of XRD peaks from distinct phases. 
Probabilistic labeling in deep learning models can be achieved by training an ensemble model or by resampling the trained model with random dropout. The probability from these methods have yet to be shown to be robust on XRD phase labeling,\citep{szymanski2021probabilistic, maffettone2021crystallography} even though there is a closed-loop experimental workflow that has already been built around it.\citep{szymanski2023adaptively} Prior work has also combined deep learning methods with differentiable physics-inspired objective functions, which force the network to factorize complex XRD spectra into physically meaningful components from a database of candidate phases,
an approach that led to the successful phase mapping of a complex ternary material system.\citep{chen2021automating} 

In this work, we propose CrystalShift, an efficient probabilistic algorithm for XRD phase labeling that complements HTE and fits well into autonomous workflows. 
It is based on a hierarchy of symmetry-constrained pseudo-refinement optimizations, best-first tree search, and Bayesian model comparison to quantify the posterior probability of potential phase combinations given a set of candidate phases.
In stark contrast with existing neural-network-based methods, the method merely needs a single spectrum to work and does not require an expensive training step based on a large number of synthetic spectra.
Further, we demonstrate that probability estimates of CrystalShift are well-calibrated, more robust against noise than existing methods, and that the algorithm exhibits higher predictive accuracy on several synthetic and experimental datasets.
In addition, CrystalShift is capable of jointly optimizing a smooth background signal during the labeling process.
The proposed method extracts rich and quantitative phase information, such as lattice strains, crystal size effect, texturing, which are critical for implementation into HTE materials discovery efforts, and provides phase combination probability estimates, which are useful both for expert evaluation and for active learning agents to model the phase space and composition-structure-property relationships more robustly, for example by quantifying uncertainty using the posterior distribution of activation probabilities.

%% file: result.tex
\begin{figure*}[t!]
  \centering
  \includegraphics[keepaspectratio, width=\textwidth]{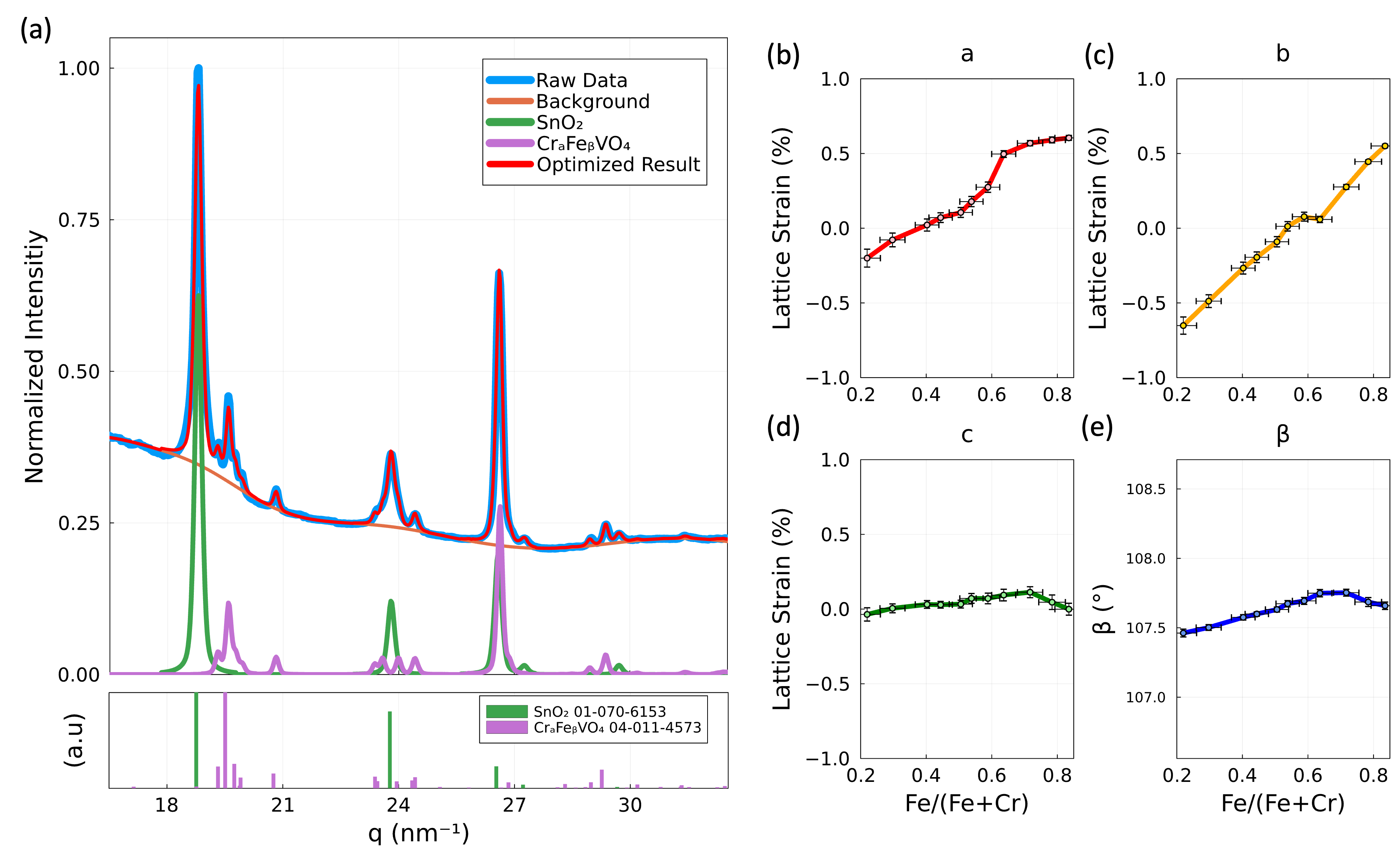}
  \caption{Result of applying CrystalShift as a pseudo-refinement method for \ce{Cr_\alpha Fe_\beta VO_4} (a) Input signal (blue) and separated signals of the background, \ce{SnO_2} and \ce{Cr_\alpha Fe_\beta VO_4} from the refinement. (b-e) The change in each of the free lattice parameters with respect to the Fe/(Fe+Cr) ratio. For a, b and c, the changes are normalized to lattice strain. For $\beta$, we show the refined angle for each composition, and the minimum and maximum angle in the figure is also set to be $\pm 1\%$ from the mean value.}
  \label{fig:refinement}
\end{figure*}
\subsection{As an Efficient Pseudo-Refinement Method}
First, we study the performance of CrystalShift by applying the pseudo-refinement optimization algorithm to 11 distinct XRD patterns collected from a sample that exhibits two-phase coexistence, namely the \ce{Cr_\alpha Fe_\alpha VO_4} monoclinic phase as a thin film on a fluorine-doped tin oxide (FTO) substrate at different Fe-Cr ratios. For this system, the monoclinic symmetry of the \ce{Cr_\alpha Fe_\beta VO_4} phase produces complex peak shifting in the XRD pattern when strained that cannot be modeled by simple multiplicative peak shifting, which is used in, e.g., convolutional NMF-based methods.\citep{suram2017automated} This multiplicative peak shifting corresponds to isotropic expansion or contraction of the lattice, a single degree of freedom that is generally unable to model alterations to the structure of non-cubic crystals.
In contrast, our method captures such behavior by modeling the diffraction peak positions using 6 or even fewer variables, depending on the crystal family of the phase. This optimization scheme properly constrains our model, forces the results to be physically sound, and exhibits a low computational complexity.

For all 11 XRD patterns associated with different Cr and Fe contents, the optimization algorithm successfully separates the constituent peaks of different phases and refines their lattice parameters accordingly. Fig. \ref{fig:refinement}a shows how the original XRD signal is correctly decomposed into the constituent patterns at a representative composition of \ce{Cr_\alpha Fe_\beta VO_4}. Since lattice parameters of the FTO substrate are known a-priori and are unlikely to shift, the regularization of such inert phases was set to be much stronger to constrain the parameters more tightly. This renders the solution of refinements much more stable, especially when we allow peak ratios to change. The XRD patterns of the \ce{Cr_\alpha Fe_\beta VO_4} phase, while having peak ratios very different from the powder diffraction pattern caused by texturing, are well-fitted, as shown in Fig~\ref{fig:refinement}a. 
The background is modeled by a kernel regressor using a Matern kernel with a long length scale and is jointly optimized with the phase model. This joint optimization prevents the background from overfitting, leading to physically meaningful decompositions of the empirical spectra.
There is some underfitting in Fig. \ref{equ:optimize}a from the background near q=18 nm$^{-1}$, which is caused by the long length scale of the kernel which constrains its flexibility. These fits demonstrate the robustness of the optimization method against non-ideal peak ratios (e.g., from texturing or atomic disorder) and complex background signals.

The refined free lattice parameters and their uncertainties, which are estimated by a scaled Hessian assuming that each of the lattice parameters are independent, are shown in Fig~\ref{fig:refinement}b for all 11 XRD samples along varying compositions. The result shows a steady increase in the a and b lattice parameters as the Fe ratio increases while the c and $\beta$ lattice parameters remain nearly constant, which is in accordance with the recently-reported manual fitting of unit cell parameters.
\citep{zhou2022high}

Based on these results, we conclude that CrystalShift is a powerful method to efficiently and accurately provide lattice information from raw XRD data of multi-phase samples to understand how the crystal structures of the constituent phases evolve under strain or as function of composition, ultimately providing valuable insight into the materials' behavior.

\subsection{Robustness of Probability Estimates}
During high-throughput XRD experiments and closed-loop autonomous experiments, sample and data quality may vary, motivating development of automated analysis algorithms the operate consistently over a range of noise levels.
\begin{figure*}[t!]
  \centering
  \includegraphics[keepaspectratio, width=\textwidth]{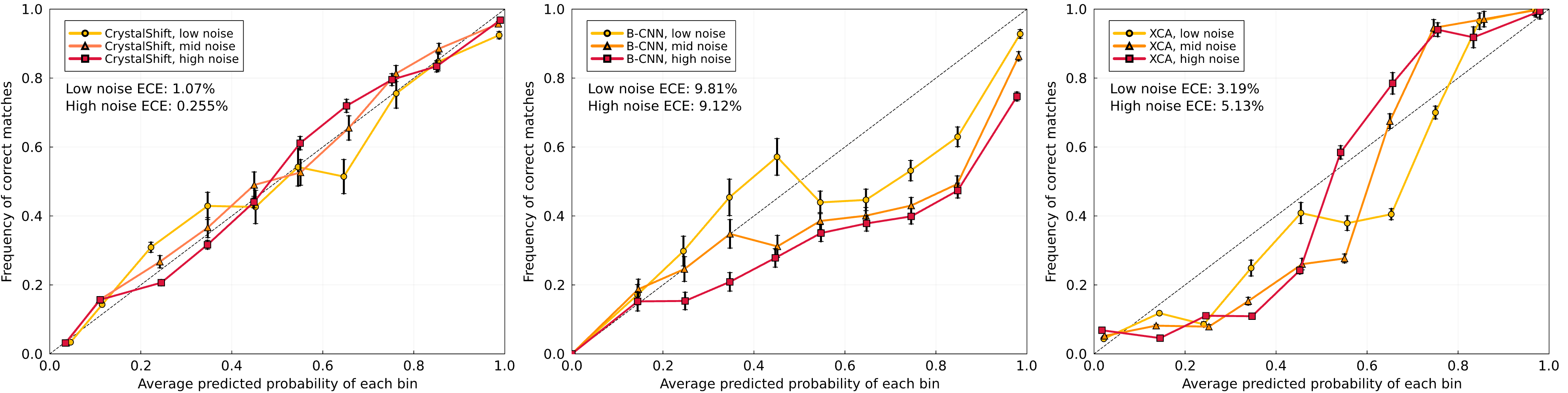}
  \caption{Probability calibration curves for (a)CrystalShift in EM mode (b) B-CNN (c) XCA with test cases that have Gaussian noise with standard deviation of 3\% (low noise), 5\% (mid noise) and 7\% (high noise) relative to the intensity of the strongest peak. The ECE values are calculated by assuming the medium noise curve of each method is perfected calibrated. Compared to other methods, CrystalShift in EM mode has 3-50 times lower ECE when the noise level deviates from when its calibrated, meaning that the probabilities it produces are much more robust against different level of noise. }
  \label{fig:calibration}
\end{figure*}
Prior work has not examined the calibration and robustness of the predicted probabilities on data with varying noise magnitudes. 
Here, we investigate how different noise amplitudes affect the quality of the probabilities of the phase combinations that are returned by CrystalShift and competing methods.
To address this question, we construct three synthetic calibration datasets with different noise levels, each containing 10,000 patterns, by randomly choosing one or two phases from five candidate phases and simulating their XRD patterns, each with random peak widths. For patterns with two phases, the phases are mixed with random fractions, predominantly in the 25-75\% range. Finally, we add noise modeled with a folded Gaussian distribution and standard deviation of 3, 5 and 7 percent of the strongest peak to the first, second, and third dataset, respectively. To evaluate the calibration of the computed probabilities, a reliability diagram is constructed for all method and noise level combinations.\citep{hamill1997reliability}

The reliability diagrams are constructed by first dividing the probability range from 0 to 1 into $M$ equally-sized bins, which are set to 10 for the results contained herein, and the prediction with probability $p_i$ belongs to the $m$th bin $B_m$ if and only if $\frac{m-1}{M} < p_i \le \frac{m}{M}$. For the $m$th bin, the frequency of correct matches, i.e. a method's accuracy in each bin, is defined as $\text{acc}_m = \sum_{i \in B_m} \mathbf{1}(\hat{y}_i=y_i)/|B_m|$ while 
the model-predicted accuracy is $\text{pred}_m=\sum_{i \in B_m} p_i/|B_m|$, where $\mathbf{1}$ is the indicator function, $\hat{y}_i$ is the predicted label of the $i$th sample, and $y_i$ is the ground truth label. 
Ideally, the average model-predicted probability should equal the empirical accuracy for each bin. In this case, the probabilities are said to be {\it calibrated}.
Formally, $\text{acc}_m=\text{pred}_m$ for all $m$, which means that the average predicted probability exactly matches the fraction of correct predictions in each bin. 
As a consequence, perfectly calibrated probabilities would give rise to a diagonal line in the reliability diagram, which we indicate by dashed black lines in Fig~\ref{fig:calibration} a-c. 
In the following, we examine how the reliability diagram for each method changes with the noise amplitude.

We show the reliability diagrams in Fig \ref{fig:calibration} by applying CrystalShift with its expectation-maximization (EM) mode (see Sec.~\ref{sec:methods} for details), together with two other neural-network-based phase identification methods that give probability estimations, the branching convolutional neural network (B-CNN)\citep{szymanski2021probabilistic} and the Crystallography Companion Agent (XCA)\citep{maffettone2021crystallography}, on the three synthetic calibration datasets. For CrystalShift, we perform the temperature scaling calibration, which is a commonly used calibration method for deep neural networks,\citep{guo2017calibration} on the 5 percent noise dataset and apply the scaling to the results of the other two noise levels. 
Note that temperature scaling is not applicable to B-CNN and XCA, as explained in Section~1 of the supplementary materials.  
Fig. \ref{fig:calibration} shows that CrystalShift gives calibrated probability estimates on datasets with varying noise levels, evidenced by the fact that its calibration lines  closely follow the diagonal line for all three noise levels. In contrast, both B-CNN and XCA have significant shifts in their calibration curves when the noise level changes and are generally not as well-calibrated as CrystalShift.

\begin{figure*}[ht]
  \centering
  \includegraphics[keepaspectratio, width=\textwidth]{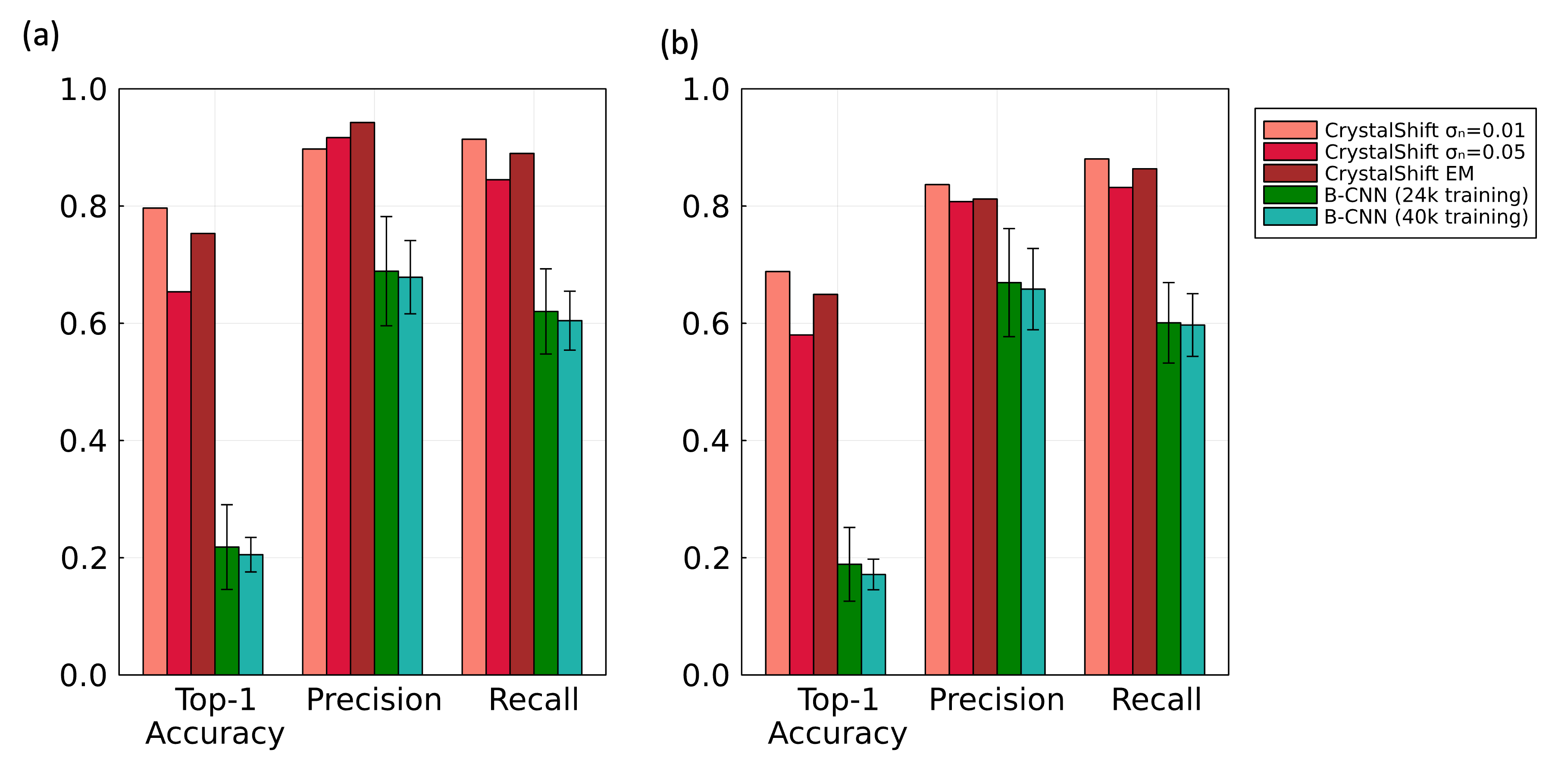}
  \caption{Top-1 Accuracy, Precision and the recall comparison of CrystalShift and B-CNN on Al-Fe-Li-O synthetic benchmark dataset, which has at most three-phase coexistence and appreciable lattice shift, with (a) 1 percent noise and (b) 5 percent noise. In both dataset, CrystalShift has higher top-1 accuracy, precision and recall while being deterministic. The drop in performance of CrystalShift at high noise variance is likely due to the fact that XRD patterns of minority phases are buried in noise. For CrystalShift, the EM method gives performance comparable to that of the manually-found optimal $\sigma_n$.}
  \label{fig:benchmark}
\end{figure*}

To fairly quantify differences in robustness against noise among the above methods, one would ideally be able to calibrate all methods, similar to the temperature scaling approach of CrystalShift. 
However, because the neural network-based methods are not directly amenable to temperature scaling, here we instead compare against an  \textit{ad hoc} transformation of the binned predicted probabilities $\{\alpha \ \text{pred}_m\}$, which improves the calibration of each binned probability $\text{pred}_m$ by multiplying it by a scaling factor $\alpha$ that is chosen to minimize the expected calibration error (ECE) for each method under 5-percent noise conditions. We subsequently apply the same scaling coefficient $\alpha$, estimated separately for each model, to the outcomes obtained at 3\% and 7\% noise levels. Finally, for each calibration curve, we calculate ECE,\citep{degroot1983comparison} a standard metric for assessing probability calibration, by
\begin{equation}
\label{eq:ece}
    \text{ECE} = \frac{1}{N} \sum_m^M |B_m| |\text{acc}_m - \text{pred}_m|
\end{equation}
which is a weighted sum of the absolute vertical distance between the points and the dashed lines in Fig \ref{fig:calibration}. The lower the ECE, the better calibrated the probabilities are. 
The calculated ECE values are listed as insets in Fig \ref{fig:calibration}. The ECE values of CrystalShift are 3-50 times lower then those of the deep learning methods. 
While additional effort work and research could likely improve the calibration of the neural-network-based methods, the results demonstrate that CrystalShift's probability estimates are already reliable and indicative of the actual misclassificaiton errors using a simple temperature scaling approach.
We attribute CrystalShift's robustness in part to the fact that it includes the standard deviation of the noise into its optimization and Bayesian model comparison steps. The explicit inclusion of the noise enables CrystalShift to disambiguate noise and phase label uncertainty. While the deep learning models are trained with noise-containing data, they do not model the noise explicitly during analysis of XRD patterns.

\begin{figure*}[ht]
  \centering
  \includegraphics[keepaspectratio, width=\textwidth]{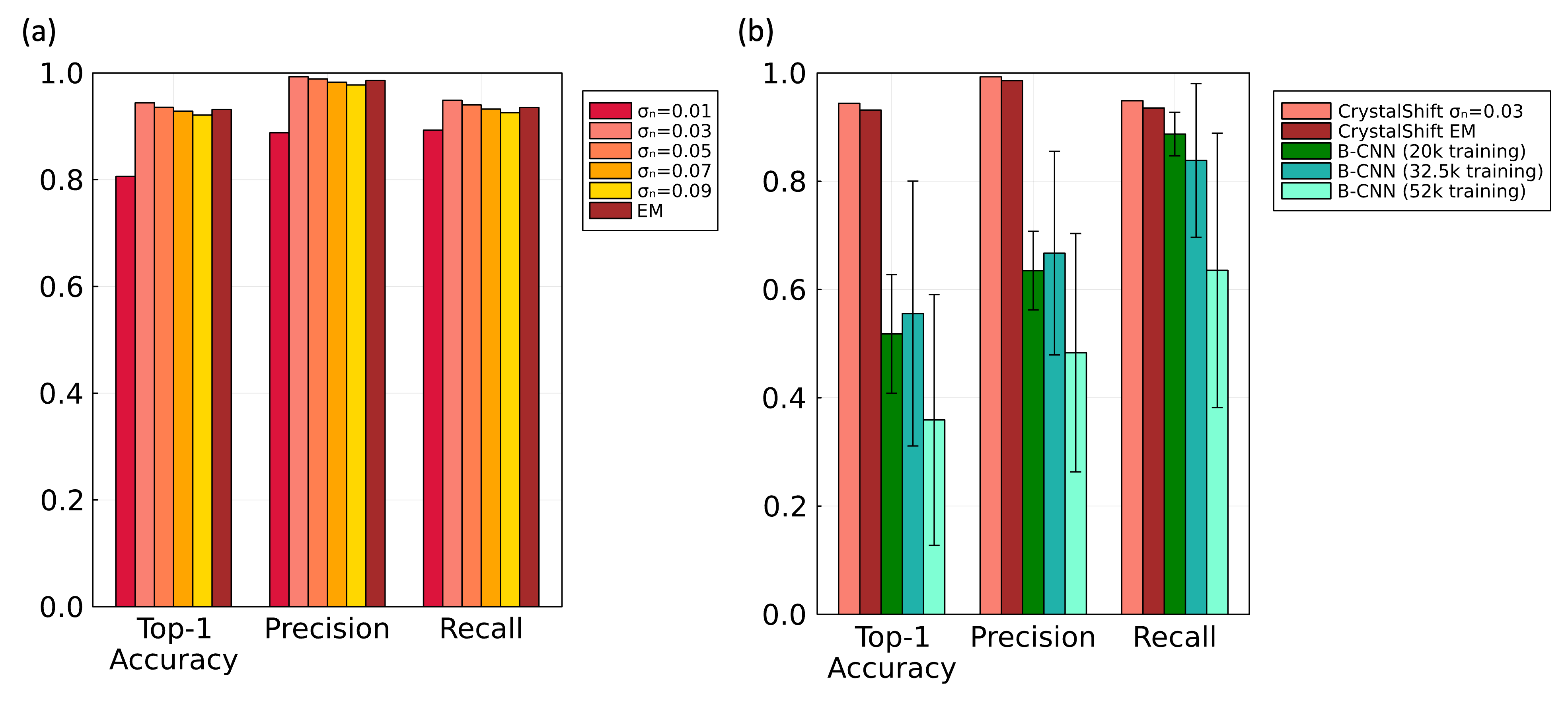}
  \caption{Top-1 Accuracy, Precision and the recall comparison of CrystalShift and B-CNN on the Ta-Sn-O experimental material system dataset after doing background subtraction with MCBL. (a) Comparing the effect of the hyperparameter $\sigma_n$ and the EM method on performance of CrystalShift. $\sigma_n$=0.03 gives the best result but EM method also perform comparitively. (b) Comparing CrystalShift performance to B-CNN with different amount of training data. Overall, CrystalShift gives better result while being deterministic.}
  \label{fig:tasno}
\end{figure*}

\subsection{Synthetic Benchmark}
To assess the performance of CrystalShift, a synthetic dataset of the Al-Li-Fe-O material system, similar to the one used in a previous paper \cite{suram2017automated}, was used to benchmark both CrystalShift and B-CNN. This dataset contains 231 patterns including 6 different phases, with each pattern being a mixture of at most 3 phases, as determined by the pseudo-ternary Al$_2$O$_3$-Li$_2$O-Fe$_2$O$_3$ oxide phase diagram. In the patterns that have phase mixtures, the phase fraction varies from 10\% to 90\% for each phase. Due to known Al-Fe site substitutions with certain phases, the XRD patterns include up to 10\% variation in lattice parameters. The width, or the standard deviation of the Gaussian, lies within 0.25 to 0.75, and was set to be higher at phase boundaries and lower when closer to the center of phase regions.

\begin{figure*}[ht]
  \centering
  \includegraphics[keepaspectratio, width=\textwidth]{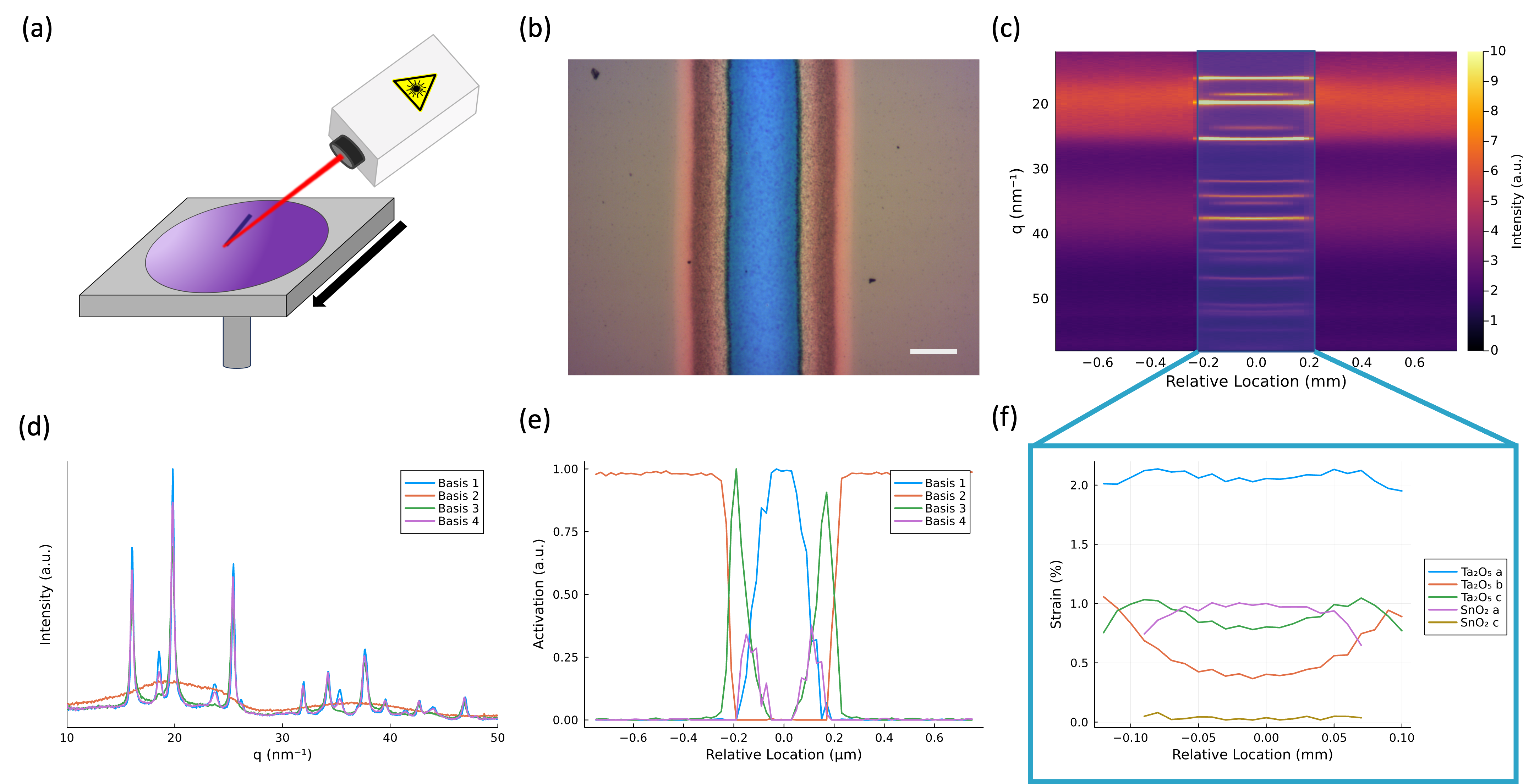}
  \caption{Application on experimental XRD patterns of LSA stripes (a) Schematic illustration of Laser Spike Annealing Setup (b) Optical microscope image of annealed stripe of a thin film. Scale bar is 200 um (c) The XRD heat map of the stripe. The onset of the strong peak signals indicate the onset of crystallization, which correspond to the position color gradient in optical image. (d) The activation (W) of the signal matrix after NMF (e) The corresponding basis (H) of the signal matrix after NMF (f) The refined lattice parameters of identified phases. }
  \label{fig:experimentapplication}
\end{figure*}

Each of the phase labelling methods is provided with 15 candidate phase prototypes and the maximum number of coexistent phases is limited to 3. We create two datasets with different noise standard deviation, 1 percent and 5 percent that of the most prominent peak, respectively. 
The benchmark metrics for each method are shown in Fig. \ref{fig:benchmark}. We benchmarked CrystalShift with two different hyperparameters and in the EM mode, and two BCNN models trained on different amount of training data. XCA was excluded because it did not provide a clear way to deal with multiphase data. We use three metrics to evaluate their performance, namely top-1 accuracy, precision, and recall. These metrics are defined as follow: the top-1 accuracy measures how often the algorithm can identify the correct phase combination as its most probable prediction in a pattern-by-pattern basis. For precision and recall, we first define the elements in the confusion matrix using only the most probable results. Specifically, we define a true positive as a phase existing in the XRD pattern and is predicted by the algorithm to be in the most probable phase combination, a false positives as a phase not being in the XRD pattern but is predicted to exist by the algorithm, and false negative as a phase that exist in the XRD pattern but is not identified by the algorithm to be in the most probable phase combination. Finally, we can calculate the precision as $\frac{tp}{tp+fp}$ and recall as $\frac{tp}{tp+fn}$, where $tp$, $fp$, and $fn$ represent the total number of true positive, false positive, and false negative, respectively.

On both noise levels, CrystalShift outperforms B-CNN on all metrics, as shown in Fig. \ref{fig:benchmark}. Additionally, increasing the amount of training data does not result in better performance of B-CNN, thus its performance is not data-limited. The large discrepancies between the top-1 accuracy and both precision and recall for B-CNN indicate that it has difficulty labeling the second or third phase.
The relatively high precision means that it can capture almost 70\% of the phases. 

For CrystalShift, the hyperparameter $\sigma_n$, which models the expected standard deviation of the noise, has a significant effect on its performance. This can be seen in the difference across all metrics in Fig. \ref{fig:benchmark}. In addition, CrystalShift does not always perform best when $\sigma_n$ equals the value of the expected standard deviation of the noise, rendering its choice challenging without comprehensive testing. We therefore propose a method to jointly optimize $\sigma_n$ using the EM algorithm and use a trick to merge it into the probability estimation framework. While the optimal value of $\sigma_n$  can outperform the EM version, the convenience of using the EM approach,  which results in a near-ideal performance, outweighs the additional cost required to optimize $\sigma_n$ in advance.

\begin{figure*}[ht]
  \centering
  \includegraphics[keepaspectratio, width=\textwidth]{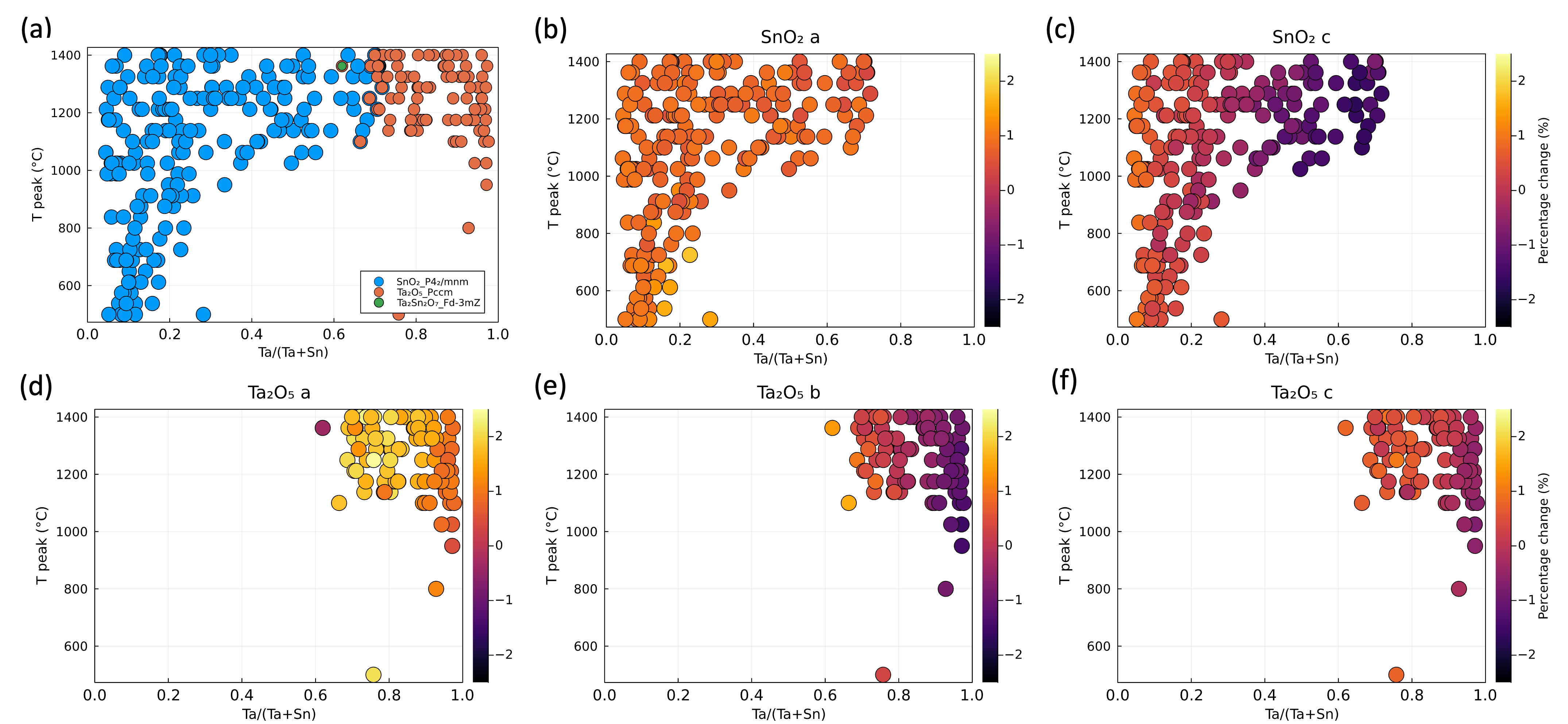}
  \caption{Analysis of \ce{Ta-Sn-O} material system. (a)Processing phase Diagram of the \ce{Ta-Sn-O} system. For ease of visualization in 2D, we disregard the dependency of the dwell time. (b-c) Changes in lattice parameters of the tetragonal \ce{SnO_2} phase. (d-f) Changes in lattice parameters of the orthorhombic \ce{Ta_2O_5} phase. Only the lattice parameters that are allowed to change are shown. The percent change is calculated using the their respective ICSD entries as references.}
  \label{fig:phasediagram}
\end{figure*}

Another difference between B-CNN and CrystalShift is that the latter is deterministic. Random dropout methods like B-CNN add randomness into the prediction results because it stochastically samples a subset of the neural network in order to estimate probabilities. In addition, for deep learning method, the training process seeks local minima of the loss function and thus can result in models trained with the same training dataset having different performances, depending on both the initialized weights and the random nature of stochastic gradient descent. In contrast, CrystalShift uses a framework based on Bayesian model comparison, which does not require training and produces deterministic labels and probabilities.\citep{mackay1992bayesian}

As the noise variance increases, both methods exhibit a drop in performance, as shown in Fig. \ref{fig:benchmark}b. Although CrystalShift still outperforms BCNN, the latter experiences a larger drop in performance. For both models, the most significant decrease is observed in the precision, indicating that there are a lot more false positives. In other words, the artificial noise are being labeled as phases. In the Al-Fe-Li-O dataset, many minority phases have fractions lower than 10\% and will be buried in the 5\% standard deviation noise. In addition, Gaussian noise may, by chance, create peak-like features that can deceive CrystalShift into mislabeling then as parts of a nonexisting phase.
For BCNN, the performance for the two dataset are almost equal, since this method always struggles to identify the minority phase regardless of what the noise level is, which is a phenomenon reported in the original paper. \cite{szymanski2021probabilistic} Because of its subtract and evaluate mechanism, the equivalent noise level becomes higher and higher for the minority phase and gives rise to inaccurate phase labels.

\subsection{Application on Experimental High-Throughput Data}
To demonstrate how CrystalShift can be applied in HTE, we apply our method to high throughput data relevant to autonomous experimentation.
Lateral gradient laser spike annealing (lg-LSA) has been shown to be a great tool for high-throughput material discovery.\citep{bell2016lateral} Its schematic is shown in \ref{fig:experimentapplication}a. A high power laser  is focused into a 1 mm by 400 um beam to rapidly heat up an oxide film deposited on top of a silicon wafer. The velocity and the large thermal mass of the stage induce high quench rate and allow this method to explore the metastable regime of material systems.

Each lg-LSA experiment creates a ``stripe'' of annealed materials, as shown in Fig \ref{fig:experimentapplication}b, wherein the composition and annealing dwell time are relatively constant, and the peak temperature of the laser annealing varies considerable as a function of the lateral position along the stripe, with the highest peak temperature being in the middle. To precisely identify phase transformations as a function of this peak temperature, and to monitor any temperature-dependent changes in lattice constants, a high density of 201 XRD patterns are acquired across each stripe using synchrotron XRD (Fig \ref{fig:experimentapplication}c). Since we anticipate that the number of very distinct XRD patterns is much smaller than the set of 201, we commence data analysis by finding representative patterns for initial analysis. Using a matrix factorization-based data reduction technique,\citep{kumar2013fast} we extract not only the 4 measured XRD patterns (the ``basis'' patterns, Fig \ref{fig:experimentapplication}d) but also the linear combination of these patterns (the ``activation'' matrix, Fig \ref{fig:experimentapplication}e) that best reconstruct the full set of 201 patterns. The 4 basis patterns may contain mixtures of phases, and some phases may appear in multiple basis patterns. Given that the probabilistic phase labelling is far more computationally expensive than that of lattice parameter refinement, this approach enables the phase labelling to be performed on only the 4 basis patterns, under the assumption that all phases present in the set of 201 patterns will appear in at least 1 of these basis patterns. \\Furthermore, once the activation of each phase in each basis pattern is determined, multiplication by the activation matrix provides a efficient estimation of the phase activation in each of the 201 XRD patterns, and the phase regions can be determined by thresholding. Each XRD pattern can then undergo the CrystalShift lattice refinement procedure with the constituent phases to provide the complete map of phase activations and lattice constants for the entire dataset. In the present case, the portion of the lg-LSA stripe that crystallized contains 2 phases, each with temperature-dependent lattice constants (Fig \ref{fig:experimentapplication}f).

Experimental XRD data has complex noise sources, including detector noise, amorphous background, air scattering, etc., and is difficult to be completely removed by current background subtraction methods. To examine how the two methods perform on experimental XRD data, the bases produced by NMF are hand-labeled by human experts to get the ground truth and those having at least one crystalline phase are background-subtracted by the multi-component background learning method and fed to the algorithms for phase labeling. \citep{ament2021autonomous} Each algorithm is given 13 candidate phases prototypes to choose from. To further study how the hyperparameter $\sigma_n$ affects the performance of CrystalShift in real application, we scan a range of different $\sigma_n$ values, shown in Fig. \ref{fig:tasno}a. We observe that the worst performance is obtained at $\sigma_n=0.01$, reaching an optimum value at $\sigma_n=0.03$, before slowly decaying at larger values of $\sigma_n$. The EM method, again, gives performance metrics very close to optimal values of each metrics. In this case, because most of the pattern are single phase, it is not surprising that the performance metrics does not drop significantly when  $\sigma_n$ increases, but the optimum can be sharper peaked for more complex materials and incentivizes the use of the EM method to jointly optimize $\sigma_n$ values.

The comparison between the two methods is shown in Fig. \ref{fig:tasno}b. Even with experimental noise, which is usually harder to deal with, CrystalShift recognizes when to ignore signals that do not correspond to crystalline phases by following Bayesian statistics, which directly models the noise and chooses the simplest model that explains the signal through Bayesian model comparison technique.\citep{mackay1992bayesian} This method results in much better accuracy, precision, and recall. The recall of B-CNN is much higher than its precision, i.e., false positives are the reason for low top-1 accuracies. This is an indication that the B-CNN method is picking up the residual experimental noise, which is inevitable even with state-of-the-art background subtraction algorithms, and label them as potential phases. This is likely due to its reliance on thresholding to terminate its branching algorithm. It is also worth noting that B-CNN has even larger error bars for experimental data than for synthetic data. It shows increasing variation between trained models with increasing training data and gives worse average performance when we increase the training data to 52000 simulated patterns, thus we attribute it to the over-fitting of training data. The deterministic nature of CrystalShift makes it preferable for high-throughput and autonomous workflows.

After using CrystalShift to label all the patterns, a processing phase diagram, which gives information regarding what phases can form at certain dwell times and peak anneal temperatures, can be easily generated by plotting the phase(s) that form(s) at the center of the stripe. The generated Ta-Sn-O processing phase diagram is shown in Fig. \ref{fig:phasediagram}a. Such automatically generated phase diagrams can significantly support expert scientists to do quickly analyze and evaluate processing phase diagrams. Taking Fig. \ref{fig:phasediagram}a as an example, information like the composition dependency of crystallization onset temperature and phase boundaries can be easily identified. Because the lattice parameters were refined during the labeling process of CrystalShift, we can also visualize how the lattice is strained to accommodate impurities, as shown in Fig \ref{fig:phasediagram}b-f for \ce{SnO_2} and \ce{Ta_2O_5}. Here, having a physical lattice model for peak shifting helps experts to understand structural changes occurring at the atomic level at various conditions. For example, the tetragonal \ce{SnO_2} has a decreasing c lattice constant with increasing \ce{Ta} content, while the a and b parameters remain almost constant. For \ce{Ta_2O_5}, we observe that only the b axis is changing when \ce{Sn} is included as impurities. This information can play an important role when creating a structure-property model on-the-fly and exploit the resulting relationship for further functional optimization. In fact, lattice distortions induced by varying composition can directly relate to changes in properties of a phase, for example in piezolectric materials. Further, the refined lattice can also be use for atomistic simulation methods like density functional theory and can be fed into AI models for additional data processing. Overall, this efficient and automated assessment of structural information from XRD and its visualization provides a powerful tool for scientist, helps bridge the gap between AI agents and human experts, and facilitates AI-human collaboration in complex material research.

%% file: discussion.tex
In summary, we have developed CrystalShift, a powerful tool for probabilistic phase labeling and rapid refinement tool, and demonstrated its utility for a range of applications where XRD data have to be automatically analyzed. Depending on the applications, the algorithm can be further refined, either to improve accuracy and precision, or to reduce compute time a for quick but less accurate analysis. In particular, CrystalShift can adapt to a given time budget by adjusting how complete the tree search is, the threshold of the refined parameters, and the desired accuracy of the probability estimates. In its current implementation, the time complexity scales with $O(nk^d)$, where n is the number of candidate phase and k is the number of top nodes that will be optimized. To improve the scaling behavior, it is possible to implement a heuristic based on matching pursuit to narrow down the candidate phase combination at each level, resulting in a time complexity of $O(n+k^d)$.\citep{tropp2007signal} This will however be the topic of future work, since our experience shows that the current implementation strike a neat balance between runtime and performance.

Despite the recent of success of deep learning methods for probabilistic phase labeling, especially the convolutional neural network lack interpretability and the convolution filters can introduce inappropriate translational invariant inductive bias for XRD patterns. The Bayesian statistical model introduced in this work circumvents this issue due to its interoperability, and the resulting simplicity and generality of CrystalShift makes it widely applicable and extendable. Though designed to label XRD patterns, CrystalShift is also applicable to any spectroscopy spectrum analysis. However, the lack of a straightforward underlying physical model for collective peak shifting in general spectra requires each peak to be optimized individually. This will greatly increase the number of parameters and result in higher computational cost. The additional degrees of freedom will also require more careful regularization to prevent overfitting. Nevertheless, a similar workflow is applicable for analyzing spectroscopy signals. If required, CrystalShift can be readily modified to model other types of spectra, only requiring a redefinition of the free variables that determine the spectrum and a method for its simulation with these variables.

Finally, CrystalShift will extend AI frameworks to pursuit increasingly complex goals, including tasks required in active learning experimentation that rely on phase information to optimally exploit model structure-to-property relations on-the-fly. The dense crystal information it extracts also make interaction between AI agents and material simulation methods like DFT attainable. Overall, the modularity and flexible of our approach will be of practical importance to establish CrystalShift as an essential tool in high-throughput material research that requires spectra demixing and labeling.

%% file: method.tex
\subsection{Crystal Modeling and Optimization}
\label{sec:crystal}
For each material system, a list of CIF files containing candidate phases has to be prepared by the user. They can be pulled from crystallography database, e.g., Inorganic Crystal Structure Database (ICSD), from atomistic simulation result, or user-defined crystal structure through softwares like VESTA or CrystalMalker. The CIFs are screened to remove duplicates so that there is only one entry for each phase.
To refine the crystal structure of the phase with respect to the given spectrum, we simulate the phase and perform optimization on the free variables of the lattice to minimize the difference between the given spectrum and the simulated one. A X-ray diffraction simulation package was used to generate the allowed $\{hkl\}$ plane indices and their corresponding peak intensities for all candidate phases.\citep{Kriegner2009} With a set of lattice parameters and a set of given plane indices $\{hkl\}$, the $q$ value of the corresponding XRD peak can be calculated by
\begin{equation}
\begin{aligned}
    p_n(\bm{\theta}_{l}) = \frac{2\pi}{V} &\Bigl(  h^2b^2c^2 \sin^2 \alpha + k^2a^2c^2 \sin^2\beta + l^2a^2b^2 \sin^2\gamma \\ 
      & + 2hkabc^2(\cos\alpha \cos \beta - \cos\gamma) \\
      & + 2kla^2bc(\cos \beta \cos \gamma - \cos\alpha) \\
      & + 2hlab^2c(\cos\gamma \cos \alpha - \cos\beta) \Bigr) ^{\frac{1}{2}}
\end{aligned}
\end{equation}
where $V$ is the unit cell volume, $a,b,c$ are lengths of the unit cell and $\alpha, \beta, \gamma$ are the unit cell angles. During the optimization, we force the phases to retain its symmetry during refinements. Therefore, the number of variables for each phase may be different, depending on their symmetries, and the vector $\bm{\theta}_{l}^{(i)}$ is a collection of the variable lattice parameters for phase $i$. The full XRD spectrum $\bm{s}$ of phase i can then be calculated by
\begin{equation}
\label{eq:phase_model}
    \bm{s}^{(i)} = \sum_{n=1} ^K I^{(i)}_n f(p_n(\bm{\theta}_{l}^{(i)}), \sigma^{(i)}, \bm{\theta}_p, \bm{q})
\end{equation}
where $I^{(i)}_n$ is the intensity of n$^{th}$ peak of the phase \textit{i}, $K$ is the number of peaks, $f$ is the peak profile function, $\bm{q}$ is the q vector of the XRD range, $\sigma^{(i)}$ is the broadening descriptor of the phase and $\bm{\theta}_p$ are the optional parameter for the peak shape. Note that $\bm{q}$ is static and is omitted in later equations for simplicity. For the peak profile function, we implemented Gaussian, Lorentzian, Pseudo-Voigt distribution. Pseudo-Voigt provides an extra parameter for tuning the peak shape and therefore is usually a good starting point. The peak shape parameter is determined by the hardware, so the parameter in Pseudo-Voigt can be fitted once and make it a fixed Pseudo-Voigt profile for optimizing other spectra to save time. If time cost is an important consideration, Lorentzian provides the fastest model evaluation while giving reasonable results.

To construct the background model, a process similar to kernel regression is used. We start by choosing a smooth kernel function, e.g. a radial basis function with defined length scale, and construct a corresponding Gram matrix with it. 
Then, the Gram matrix is decomposed using the singular value decomposition (SVD) to extract the most important components of the matrix, only keeping the bases with singular values larger than a set threshold value to reduce the computational cost of the model.
We then use the linear combination of eigenvectors to model the background as
\begin{equation}
    \bm{b} = \bm{U} \bm{\theta_b}
\end{equation}
where $\bm{U}$ is the orthonormal basis from the SVD of the Gram matrix and $\bm{\theta_b}$ is the coefficient for each basis.

There are two version of optimization for different applications.
For efficient probabilistic phase labeling, peak intensities are constrained to be constant, i.e. equal to the values of the original candidate structures.
For each phase combination, the method then solves the following optimization problem
\begin{equation}
\begin{aligned}
\label{equ:optimize}
    \argmin_{\bm{\theta}_{l_i},c_i,\sigma_i, \bm{\theta}_p, \bm{\theta_b}} || & (\sum_{i=1}^P c_i\bm{s}^{(i)}(\bm{\theta}_{l}^{(i)}, \sigma^{(i)}, \bm{\theta}_p) \\
    &+ \bm{U} \bm{\theta_b} - \bm{t})/\sqrt{2}\sigma_n ||_2^2 + r^2,
\end{aligned}
\end{equation}
where $c_i$ is the activation or fraction of each of the constituent phase, $\bm{t}$ is the experimentally observed XRD spectrum and $r^2$ is a regularization term. Note that the background model can optionally be included. 
The regularization $r^2$ has the following form:
\begin{equation}
    \label{equ:regularize}
    r^2 = \sum_{i=1}^P \Big|\Big|  \frac{\bm{\theta}_{l}^{(i)}-\bm{\theta}_{r}^{(i)}}{\sigma_{r}^{(i)}} \Big|\Big|_2^2 + \lambda||\bm{\theta}_b \oslash \text{diag}(\boldsymbol{\sigma}_b)||_2^2
\end{equation} 
where $\bm{\theta}_{r}^{(i)}$, ${\sigma}_{r}^{(i)}$ and $\lambda$ are user-defined penalizing weight, $\boldsymbol{\sigma}_b$ are the singular values of the Gram matrix that defines the background model, and $\oslash$ indicates element-wise division. $\bm{\theta}_{r}^{(i)}$ are generally set to the lattice parameters in the CIFs and ${\sigma}_{r}^{(i)}$ determines how strongly the user wants to penalize lattice distortion during the optimization.
$\lambda$ is usually set to 100 to keep a complex background model from absorbing non-background peaks. This optimization is carried out until it converges or after it has run through a fixed number of iterations. The optimization method makes use of \texttt{ForwardDiff.jl}, an auto-differentiation package to obtain the Jacobian and Hessian of the loss with respect to the parameter vector.\citep{RevelsLubinPapamarkou2016}

We used an expectation maximization (EM) method to jointly optimize the hyperparameter $\sigma_n$ and the other parameters of the model.\citep{dempster1977maximum}
The EM algorithm starts with an initial estimate of $\sigma_n$ and proceeds by carrying out the optimization step described in eq. \ref{equ:optimize}, which constitutes the maximization step. 
Afterward, the expectation step computes the standard deviation of the residual after the maximization step and assigns it to $\sigma_n$. 
This process is then alternated until convergence or a pre-specified number of iterations.

For the optimization process that targets cell-refinement, the peak intensities are also optimized and a block coordinate descent strategy is deployed.\citep{ortega2000iterative}
The algorithm alternates between optimizing peak intensities and optimizing other parameters.
For a given phase combination, the same optimization problem as in Eq.~\ref{equ:optimize} is solved for a number of optimization steps. 
Then, the algorithm switches to optimizing the peak intensities of the phases $I_n$ while keeping all other parameters fixed:
\begin{equation}
\label{equ:peak_opt}
     \argmin_{\bm{I}^{(1)},...,\bm{I}^{(P)}} || \sum_{i=1}^P c_i\bm{s}^{(i)}(\bm{I}^{(i)}) + \bm{b} - \bm{t} ||_2^2 + \ell
\end{equation}
where $\ell$ is the regularization for peak intensity modification
\begin{equation}
    \ell = \sum_{i=1}^P \Big|\Big|  \frac{\bm{I}^{(i)}-\bm{I}^{(i)}_{r}}{\sigma_{I}} \Big|\Big|_2^2
\end{equation}
where $\bm{I}_{r_i}$ and $\sigma_{I}$ are user-defined regularization parameters. Typically, $\bm{I}_{r_i}$ is set to the intensity from the XRD simulation package.
Both eq. \ref{equ:optimize} and eq. \ref{equ:peak_opt} can be solved by common numerical optimization methods.
However, the Levenberg-Marquart (LM) algorithm\citep{levenberg1944method,marquardt1963algorithm} in particular shows good balance between convergence speed and execution time, and arrives at a good minimum for solving Eq.~\ref{equ:optimize}.
Notably, BFGS achieves better results optimizing Eq.~\ref{equ:peak_opt} empirically.\citep{fletcher2000practical}
Although the exposition here focuses on the $\ell_2$-objective, it is also possible to carry out the optimization with respect to other objective functions, such as the Kullback–Leibler divergence or the Jensen–Shannon divergence. However, this would require selecting a different optimization method because LM is limited to quadratic losses. Note that the parameter optimization is performed in log space to constrain all the parameter to be positive.

\subsection{Best-First Tree Search}
To search the combinatorial space of possible phase combinations efficiently, a tree structure that incorporates the greedy heuristic of forward stepwise regression\citep{miller2002subset} was constructed as follows: 
First, an empty root node is constructed. To grow the tree into the next level, the method generates a series of descendent nodes each containing both the phases in its parent node and one additional phase that is not in its parent node. We call this process ``expand''. The optimization method described in Section~\ref{sec:crystal} is then applied to optimize all the nodes in this level. 
When constructing the next level of the tree, only $k$ nodes that have the lowest residual are expanded.
This is the greedy selection criterion referenced earlier.
This process is repeated until the desired level, i.e. the maximum allowed number of coexisting phase, is reached. In this way, the tree is constructed dynamically and lazily, which means that nodes are only constructed when they are needed so that the combinatorial number of phase combinations do not have to be enumerated explicitly. Note that all optimizations at the same level are independent and can therefore be parallelized easily.

\subsection{Probabilistic Modeling}
In order to quantify the probability that a given phase combination is present in an observed spectrum, CrystalShift uses Bayesian model comparison techniques, which require access to the Bayesian model ``evidence'' , a quantity that is computed by marginalizing (i.e. integrating) out the model's parameters weighed by their prior probabilities.
Unfortunately, this cannot be done analytically for the non-linear phase model in Eq.~\eqref{eq:phase_model}. 
For this reason, we use a Laplace approximation for each of the tree search results to approximate the marginal likelihood.\citep{mackay2003information} 
To this end, we define the log likelihood of 
any model $M^{(i)}$ -- i.e. a combination of phases and a node in the tree --
 to be
\begin{equation}
    \log p(\mathbf{t}|M^{(i)}, \mathbf{\theta}^{(i)}) = -L_i.
\end{equation}
Note that the quadratic terms in Eq.~\ref{equ:optimize} are proportional to Gaussian probability densities upon exponentiation.
When using the EM algorithm, we set $\sigma_n$ to be the lowest EM-inferred $\sigma_n$ of all nodes in the tree before estimating the marginal likelihood,
because the minimum value is more likely to reflect the true $\sigma_n$ of the data and exceedingly unlikely phase combinations can give rise to outliers in the EM-inferred $\sigma_n$ values.

CrystalShift then employs a Laplace approximation,
which can give accurate estimates of the marginal likelihood without the use of computationally expensive Monte-Carlo sampling,
if the likelihood as a function of the variables is well-peaked. 
In particular, the Laplace approximation of the marginal likelihood is given by
\begin{equation}
    \log p(\mathbf{t}|M^{(i)}) \approx -L_i + \frac{\log|H_\mathbf{\theta}^{-1}(L)|}{2} + d\frac{\log(2\pi)}{2},
\end{equation}
where $H_\mathbf{\theta}$ is the Hessian of the loss with respect to the variables and $d$ is the number of variables at the maximum a-posteriori (MAP) estimate
of $\theta$.
Notably, this approach primarily requires 1) the computation of the MAP parameters $\theta$ via numerical optimization, and 2) the computation of the Hessian at the MAP parameters, which is computationally cheap via forward-differentiation\citep{RevelsLubinPapamarkou2016}  because the number of variables governing an XRD pattern is usually small. 
The posterior model probabilities can then be computed by normalizing the marginal likelihoods of all considered models:
\begin{equation}
\label{eq:posterior_model_probs}
    p(\mathbf{t}|M^{(i)}) = \frac{p(\mathbf{t}|M^{(i)})} {\sum_{j=1}^N p(\mathbf{t}|M^{(j)})}
\end{equation}
The probability can further be calibrated by running a set of simulated noisy spectra through the labeling process to create a reliability diagram and minimizing the ECE, as defined in Eq~\eqref{eq:ece}. 
Empirically, low ECE can be achieved with the temperature scaling method, which scales the log marginal likelihoods by a scalar factor before exponentiating and normalizing them to attain the posterior probabilities as in Eq.~\eqref{eq:posterior_model_probs}.\citep{guo2017calibration}

\subsection{Library Synthesis}
\paragraph{TaSnO sample preparation}
Thin film is deposited on top of a heavily-doped silicon wafer (0.01-0.02 $\Omega \text{-cm}^{-2}$) using a AJA radio-frequency (RF) sputtering system, in which has three target guns to enable composition gradient depositions. The composition-location relation was measured by x-ray fluorescence (XRF) and showed that the cation ratio spans across a wide range, as shown in Fig. \ref{fig:phasediagram}. The thin film was then processed by lg-LSA \citep{bell2016lateral} with a condition grid that covers the anneal dwell times from 250 $\mu$s to 10 ms and peak temperatures from 400 to 1400 $^\circ$C.

\paragraph{\ce{CrFeVO} sample preparation}
The Cr-Fe-V oxide composition library was synthesized by reactive co-sputtering of metal targets in a custom-designed combinatorial sputtering system \citep{suram2015combinatorial} at room temperature and followed by post-deposition anneal in a box oven in flowing air at 650 °C for 1 h. The library was deposited in a mixed atmosphere of \ce{O_2} (0.6 mTorr) and Ar (5.4 mTorr) using Cr, Fe, and V sources placed 120° apart with respect to the substrate plane of the 100-mm-diameter Pyrex glass with a conductive \ce{SnO_2}:F coating. The deposition proceeded for 10 h with the RF powers on Cr, Fe, and V sources set to 42, 62, and 150 W, respectively, to obtain the desired composition spread for which oxygen stoichiometry was not specifically controlled. The bulk metal compositions were characterized by XRF measurements using an EDAX Orbis Micro-XRF system with an x-ray beam of 2 mm in diameter. The Cr K, Fe K, and V K XRF peak intensities were extracted from the Orbis software and converted to normalized cation compositions using the sensitivity factor for each element calibrated by commercial XRF calibration standards (MicromatterTM).

\subsection{XRD data Collection}
\paragraph{CrFeVO XRD collection}
The bulk crystal structure and phase distribution of Cr-Fe-V oxide composition library was determined by XRD measurements. XRD was acquired using a custom high throughput setup\citep{gregoire2014high} incorporated into the bending-magnet beamline 1-5 of the Stanford Synchrotron Radiation Light Source (SSRL) at SLAC National Accelerator Laboratory. The characterization employed a monochromated 12.7 keV source in reflection scattering geometry with a 2D image detector. Diffraction images were processed into one-dimensional XRD patterns using WxDiff software with calibration from a \ce{LaB_6} powder standard, and further analyzed in the Bruker EVA software.

\paragraph{TaSnO XRD collection}
The Ta-Sn-O dataset was collected at Cornell High Energy Synchrotron Source (CHESS) ID3B beamline. Compound reflective lens focuses the 9.7keV X-ray beam into a 20$\times$90 um spot with a 2$^\circ$ incident angle. A Dectris Eiger 1M detector was used to collect the diffraction signal. When collecting the data of an annealed stripe, the precision stage that carries the wafer slowly move horizontally and let the X-ray beam scan across the stripe while the detector collect image with a constant exposure time. 201 XRD patterns at different horizontal positions were collected per anneal stripe at a 10 $\mu$m interval. PyFAI was used to integrate the 2d XRD pattern from the detector into 1d XRD pattern in q space. \citep{ashiotis2015fast}